\documentclass[twocolumn,aps,floats,floatfix,superscriptaddress]{revtex4}
\usepackage{graphicx,amssymb}
\usepackage{epsfig}

\newcommand{\beq}{\begin{equation}}
\newcommand{\eeq}{\end{equation}}

\newcommand{\bX}{{\bf X}}
\newcommand{\bY}{{\bf Y}}

\begin{document}
\draft
\title{Generalization of conformal mapping to scattering of electromagnetic waves from surfaces: An example of a
triangle}
\author{ S. T. Chui}
\affiliation{ Bartol Research Institute and Dept. of Physics and Astronomy, 
University of Delaware, Newark, DE
19716, USA}
\author{ Shubo Wang}
\affiliation{ 
Department of Physics, The Hong Kong University of Science and Technology, Clear Water Bay, Kowloon, Hong Kong, China}
\author{C. T. Chan}
\affiliation{ Department of Physics, The Hong Kong University of Science and Technology, Clear Water Bay, Kowloon, Hong Kong, China}
\begin{abstract}
We discuss a way to exploit the conformal mapping to study the response of a finite metallic element of arbitrary shape to an external electromagnetic field at finite frequencies.
This provides a simple way to study different physics issues  and provides new insights that include  the issue of vorticity and eddy current, and the nature of the divergent electric field at the boundaries and at corners. The nature of the resonance can be directly addressed and clarified. 
We study an example of an equilateral triangle and found good agreement  with results obtained with traditional numerical techniques.

\end{abstract}
\maketitle
\section{Introduction}

In electrostatics in two dimensions, conformal mapping has provided much analytic understanding. This technique,
though powerful, is limited to two dimensions and only to problems in electrostatics.  
The calculation of conformal harmonic maps has seen much progress recently
in computational conformal geometry\cite{bookm}.  For example, any surface of genus 0 and 1 can be mapped into a finite disk and an annulus with harmonic conformal maps, the numerical implementation of which has already been accomplished\cite{bookm}. 
In this paper we propose a rigorous and numerically efficient way to exploit conformal mapping to deal with the finite frequency electromagnetics problem. This follows
our recent work that generalized our 
circuit approach for wires\cite{book} to finite disks and annuli\cite{disk}. When tested for arrays of split rings, our approach is found to be two orders of magnitude faster than COMSOL\cite{zhan}.  Our approach provides new insight on the nature of the divergence of the field at corners, the systematic study of resonances, and the emergence of eddy currents caused by corners. 
We illustrate our results with the example of a triangle.

The corners of a metallic plate is expected to exhibit a divergent field under external excitation. This is difficult to treat and understand with approaches involving a finite mesh. As the corner is approached, the mesh size has to become smaller and smaller. In the study of metamaterials, the exploitation of the strong fields at the corners have been discussed, but the determination of these fields is numerically challenging and the physics of their effect on the resonance is not well understood. Our approach clarifies the nature of this effect. We find that the electric field is a product of two factors. At a resonance, the first factor represents the field that diverges at {\bf all} edges; the second factor represents the divergence that is more severe but integrable at the corners. The integrable singularity at the corners is built in through the conformal mapping and there is no numerical difficulty in dealing with the corners.

The term "eddy current" is often used to describe the interaction of electromagnetic waves with metallic surfaces.  
In our approach, local eddies with finite local vorticity comes naturally into the description of resonances that we find through the conformal mapping. A current going into a corner has to turn around and an eddy is formed. This local vorticity of a resonance exists even though its average, the total magnetic dipole moment, can be zero. 

Much of the interesting physics of the response of finite structures is contained in the nature of its resonances. Usually this is extracted from the numerical solution for the scattering properties. It is difficult to obtain a clear physical understanding and extract information of the resonances with this approach, however. The extraction is cumbersome and the calculation is numerically demanding. Such information are usually incomplete. For example, information on dark or grey resonances not strongly coupled to particular external excitations cannot be easily obtained. The possible degeneracy of the resonances is not manifested, and the effect of the background has to be eliminated.  Our method provides a direct and physically simple way to explore this information. We study the resonances of a equilateral triangular plate and find three kinds of low-lying resonances (a) Regular doubly degenerate resonances with a finite electric susceptibility,  local vorticity but no global magnetic dipole moment.  
(b) Grey doubly degenerate resonances with an order of magnitude smaller electric susceptibilities, local vorticities but no global magnetic dipole moments (c)  nondegenerate electrically dark resonances with  finite magnetic dipole moments and no electric dipole moments.

This paper is organized as follows. We develop our theory in section II. The calculation of the
circuit elements are discussed in section III. Numerical results are presented in section IV and compared with numerical solutions of this problem with other traditional methods.

\section{Circuit theory}

We are interested in the current flow on a finite surface caused by an external electromagnetic (EM) wave. 
An example of such a structure  is the equilateral triangle.
We assume that the film is thin enough that there is no current in the direction perpendicular to it. The current density ${\bf j}$ 
in the presence of an external electric field $E_{ext}$ is governed by the equation
\beq
\rho{\bf j+ E_{em}=E}_{ext}.
\eeq
where $\rho$ is the resistivity, ${\bf E}_{em}$ is the electromagnetic (EM) field generated by the current, ${\bf E}_{ext}$ is the external EM field.
We impose the boundary condition of no current flow perpendicular to the boundary of the film  with a large
boundary resistivity $\rho_s$ which we take to approach infinity\cite{book,disk}. The total resistivity $\rho$ is a sum of this surface term and a metal resistivity $\rho_0$.  
The electric field $E_{em}$ can be obtained from the integral form of Maxwell's equation . It is a sum of a capacitive and an inductive term: $E_{em}=E_c+E_L$. These terms can  be expressed in terms of the current density as    
\beq
E_C=-i/(\omega\epsilon_0)\int d{\bf r}' G_0(r-r') \nabla'\nabla' \cdot j(r^{\prime}) . 
\eeq 
\[
E_L= -i\mu_0\omega\int  d{\bf r}' j(r^{\prime})G_0(r-r'). 
\]
where the bare Green's function $G_0$ is given by
\beq
G_0=\exp(ik_0|r-r'|)/4\pi|r-r'|.
\eeq

In our approach, we represent the currents and the fields of interest {\bf not} in terms of finite elements on a mesh but in terms of a set of orthonormal basis functions. As we discuss before\cite{book,disk}, the impedance matrix  becomes nearly diagonal and the convergence is very fast. Usually this makes the implementation of the boundary condition difficult but, as we explain below, there is a simple way to implement this condition in terms of extra parameters, similar to the idea of the Lagrange multiplier. For very simple cases, the basis functions are the well known special functions. For a given finite surface, there is a harmonic conformal map that maps it into a disk.  We construct the basis function for this surface from the basis function of a circular disk with the conformal harmonic map.  

The electromagnetic field $E_{em}$ can be represented as
${\bf E}_{em}={\bf Z^0 j}$ where the
"impedance" matrix $\textbf{Z}^0$
is just the the representation of the Green's function in this basis.
More specifically, 
\beq
\textbf{Z}^0=-i\omega \textbf{L} +ic^2/(\omega \textbf{C})
\eeq
where, for any basis function $\bX$, $\bY$
\beq
L_{Xi,Yj}=\mu_0\int d{\bf r} d{\bf r'} [\bX_i(r)]^*\cdot \bY_j(r')G_0(r,r')
\eeq
\beq
(1/C)_{Xi,Yj}=\int d{\bf r} d{\bf r'} [\bX_i(r)]^*\cdot\nabla'\nabla'\cdot \bY_j(r')G_0(r,r').
\eeq
As we shall see, when the basis functions are orthonormal, the off-diagonal elements of the impedance matrix is much less than the diagonal one\cite{book,disk}. Furthermore the magnitude of the impedance increases rapidly. These greatly  facilitate the convergence of the solution and provide for a much better understanding of the physics. We next discuss the construction of the new basis states.

\section{The basis functions}
We first recapitulate the basis function for a circular disk, which is given by the usual vector cylindrical functions:
$${\bf M}_n({\bf r})=( f_{n+1}\textbf{e}_-+ f_{n-1}\textbf{e}_+)/2;$$ 
$${\bf N}_n({\bf r})= (-f_{n+1}\textbf{e}_-+f_{n-1}\textbf{e}_+)/2.$$ 
${\bf e}_{\pm}={\bf e}_x\pm i {\bf e}_y)$. For fields in a closed compact region including the origin 
$f_{n\pm 1}({\bf r})=\exp(i(n\pm 1)\phi) J_{n\pm 1}(x)/C_n.$ $J_n$ is the Bessel function, $x=k_{\perp}r.$  The normalization
constant is given by
$C_n^2=(C_{0,n+1}^2+C_{0,n-1}^2)/2$ where $ C_{0mk}^2/(2\pi)=\int J^2 rdr=
J^2(R^2-m^2/k^2)/2+R^2J^{\prime 2}/2.$ 
\textbf{M}, \textbf{N} can also be written as
$${\bf M}_n= \exp(in\phi) [inJ_n(x)/x{\bf e}_r-J_n'(x){\bf e}_{\phi}],$$
$${\bf N}_n=\exp(in\phi) [iJ_n'(x){\bf e}_r- nJ_n(x)/x{\bf e}_{\phi}]$$
where ${\bf e}_r,$ ${\bf e}_{\phi}$ are unit vectors in the radial and azimuthal directions respectively.
A boundary condition  can be imposed on the basis functions.
There are two possibilities. the wave vectors k can be chosen so that 
for the "D" (for "derivative") basis,  $J_m'(k_{Dm}R)=0$
whereas for the "V" (for "value")  basis,  $J_m(k_{Vm}R)=0.$ For either case the $k_{m\perp}$s
form a discrete spectrum.

We next turn our attention to constructing the basis functions for a non-circular finite surface. We assume that there is a conformal harmonic mapping 
$w=u+iv=w(x+iy)$ between a point $w$ in a circle and a point $z=x+iy$ in the finite surface.  An example of such a conformal mapping to the equilateral triangle is recapitulated in Appendix 1. 
We shall expand the vector fields in terms of the  new basis functions $c{\bf M}(w(z))$, $c{\bf N}(w(z))$ with this mapping\cite{detail2} :
Here $$c{\bf M}(w(z))=( g_{n+1}\textbf{e}_-+ g_{n-1}\textbf{e}_+)/2;$$ 
$$c{\bf N}(w(z))= (-g_{n+1}\textbf{e}_-+g_{n-1}\textbf{e}_+)/2.$$ 
where
\beq
g_{n\pm 1,k'}=p_{\pm}f_{n\pm 1}.
\eeq
We have picked $p_+=dw^*/dz^*=1/(dz^*/dw^*),$ $p_-=dw/dz$. 
For the equilateral triangle, $dz/dw=(1-w^3)^{-2/3}.$
With this choice, the new basis functions are orthonormal with the corresponding measure: 
$\int dz dz^* {\bf (cM_n)^*\cdot cN_m}=
\int d^2w \rho^{-1} |p|^2 {\bf M_n\cdot N_m}=0$ where $\rho=|dw/dz|^2$ is the Jacobian.  

In two dimensional electrostatics,  analytic functions $w(z),\  w^*(z^*)$ satisfies the Laplace equation.
The  corresponding electric fields are given by 
$\nabla w={\bf e}_+(dw/dz)={\bf e}_+p_-,$ $\nabla w^*={\bf e}_-p_+.$
These are equal to the long wavelength limit of our basis functions  
${\bf e}_+g_{1-1,k'}(w),$ ${\bf e}_-g_{-1+1,k'}(w).$
This is one way that we can think of our approach as a generalization of the method of conformal mapping from electrostatics to finite frequencies. The integrable divergence of the field at the boundary is explicitly 
incorporated in our solution.  

\subsection{Electric moment}
Our choice of basis functions is motivated by the simplification in the algebra of the derivatives. We find (See Appendix 2) that with respect to the coordinates ${\bf r}=(x,y)$ for the finite surface,
\begin{equation}
\nabla\cdot c{\bf M}=0.
\end{equation}
\begin{equation}
\nabla \cdot c{\bf N}=-k f_n|dw/dz|^2.
\label{div}
\end{equation} 
The states  c{\bf M} has zero divergence.
 From the continuity equation the charge density is related to the divergence of the current density: $n=\nabla\cdot j/(i\omega).$ Because $\nabla \cdot c{\bf M}_m=0,$ $c{\bf M}_m$ have no electric dipole moment.  A current distribution described by this state is of magnetic character and does not generate a finite charge density. The electric dipole moment of a current density proportional to ${\bf N}_1$ is given by 
\beq
{\bf P}=-ik/\omega \int d^2r \exp(i\phi)J_1(kr){\bf r}/C_{1N}
\label{d}
\eeq
where the range of integration is within the unit circle. The Jacobian is cancelled out.
$c{\bf N}_m$ has an $m^{th}$ order electric multiple moment.
These integrals can be carried numerically from the series expansion of the Bessel function: 
$\int_0^k dx x^2J_n(x)/k^3=\sum_{r=0} (-1)^r(k/2)^{n+2r}/r!/(m+r)!/(n+2r+3).$

\section{Magnetic Field}
\subsection{Local Vorticity}
A new physics phenomenon for our system is the development of local circulation near the sharp corners.
Our choice of the basis functions also incorporates the fact the vorticities and eddies are developed at corners where the current has to turn around. The system can be {\bf locally} magnetoelectric. 
Physically, if a current rush at a sharp corner there can be a charge pileup. To take care of this, the current develop local eddies and turn around near the corner. This local eddy is manifested as a local magnetic field. Locally, inside the triangle, the electric field is just ${\bf E=\rho_0 j}$. The magnetic field is given by
$i\omega {\bf B=\nabla\times E}.$ 

We (See Appendix 2) find that 
\beq
({\bf \nabla\times M})_{nz} =i k|dw/dz|^2 f_{n},
\label{b}
\eeq
$$({\bf \nabla\times N})_z=0.$$
Only the basis functions c{\bf M} are locally magnetic. 
Eq. (\ref{b}) implies that there are two contributions to
the vorticity; a "global" contribution from the factor $f_n$ and a "local" contribution from the factor
$|dw/dz|^2$. This local contribution comes from the effect of the shape and is not present for a simple disk. 

\begin{figure}[tbph]
\vspace*{0pt} \centerline{\includegraphics[angle=0,width=6cm]{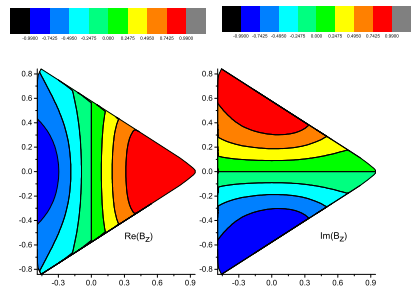}} 
\vspace*{0pt}
\caption{ The real and imaginary parts of the ${\bf B_z=\nabla\times M}^D(l=1,k=1).$ The length units are such that the distance from the center to the corner is 1.}
\label{bd1m}
\end{figure}

\begin{figure}[tbph]
\vspace*{0pt} \centerline{\includegraphics[angle=0,width=6cm]{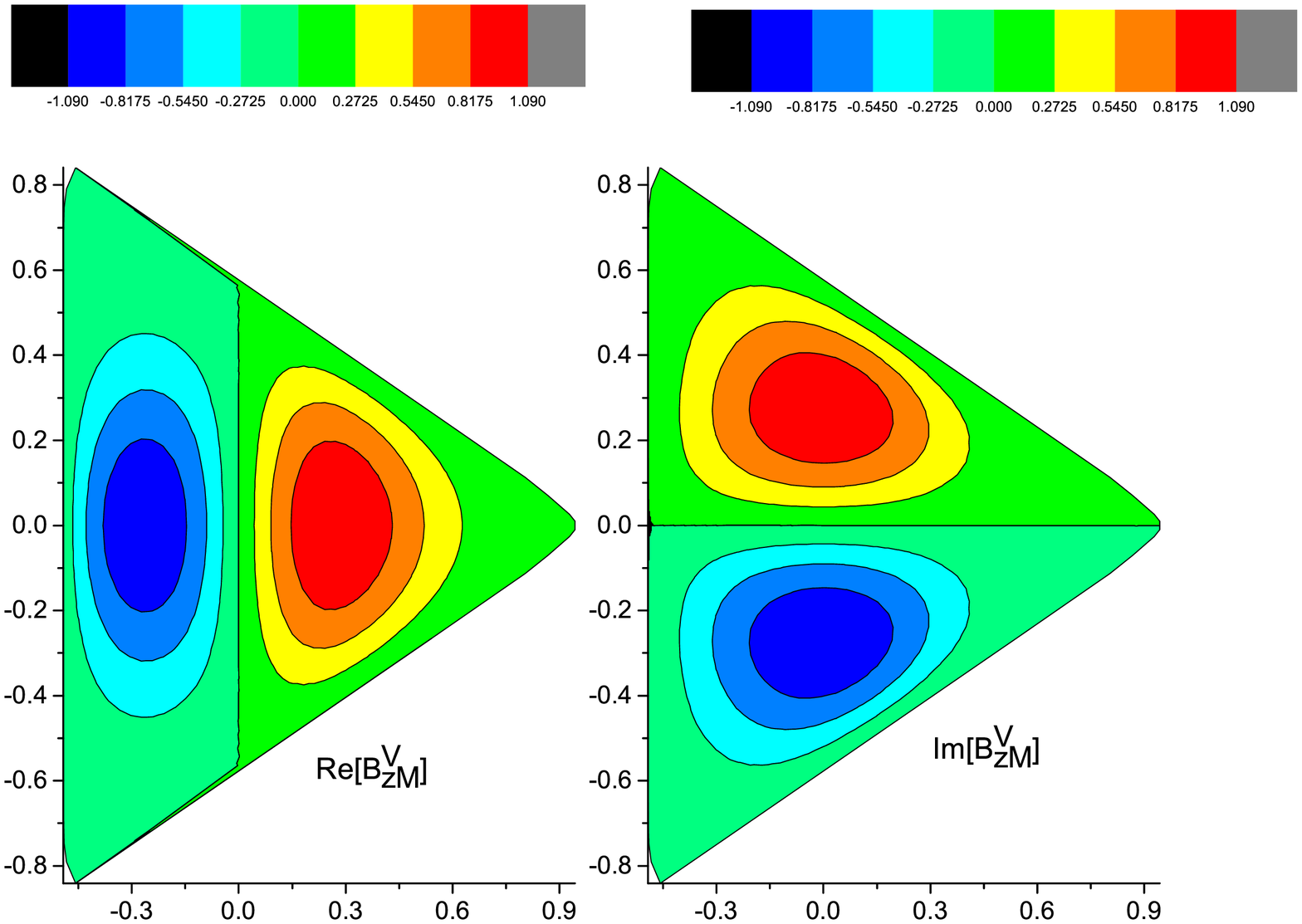}} 
\vspace*{0pt}
\caption{ The real and imaginary parts of the ${\bf B=\nabla\times M}^V(l=1,k=1).$ The length units are such that the distance from the center to the corner is 1.}
\label{bv1m}
\end{figure}

\begin{figure}[tbph]
\vspace*{0pt} \centerline{\includegraphics[angle=0,width=6cm]{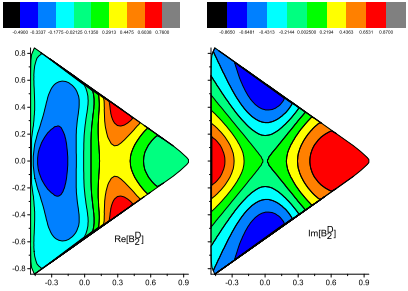}} 
\vspace*{0pt}
\caption{ The real and imaginary parts of the ${\bf B= \nabla\times M}^D(l=2,k=1).$ The length units are such that the distance from the center to the corner is 1.}
\label{b2dm}
\end{figure}

In Fig. (\ref{bd1m}-\ref{b2dm}) we show the local vorticity (the magnetic field) of some of our basis functions. This local vorticity are of opposite directions at different places of the figure. As can be seen 
from the formula for the divergence and the curl, ${\bf M}$ 
corresponds to circular flow with fractional vortices whereas ${\bf N}$ exhibits sources and sinks that corresponds to a finite divergence. For the V type boundary conditions, the "vortex core" occurs inside the sample and not at the edges.

\subsection{Global magnetic moment}
We can also look at the vorticity from a global view.
Because the total current may not be zero at any time, the magnetic moment can depend on the choice of origin.
The total magnetic dipole moment with respect to the origin  is given by $\Gamma=\int (xj_y-yj_x).$ The current is written as 
${\bf j}=\sum_{i=\pm}j_{-i}{\bf e}_i$ . 
$\Gamma
=i\int [-j_+z^*+j_-z]$.
For the basis functions {\bf cM} ({\bf cN}), $j_-= + (-) J_{n- 1} \exp[i(n-1)\phi] p_{-}/2,$
$j_+= J_{n+ 1} \exp[i(n+1)\phi] p_{+}/2.$
Thus 
\beq
\Gamma_n=i\int_{\triangle} r\exp(in\phi)[-J_{n+1}p_+\pm J_{n-1}p_-].
\eeq
where the symbol $\triangle$ indicates that region of integration is within the finite figure.
Transformation the region of integration to within the unit circle and cancelling out the Jacobian 
factor, we find that $\Gamma_n=0$ except for n=0, with 
$$
\Gamma_0=-i\int r J_1 [1/p_-\pm 1/p_+].
$$

\section{Circuit Equations}
The current in the plane can  be expanded in terms of our basis as
\beq
{\bf j}=\sum j_{Mm}(k)c{\bf M}_m(kr)+j_{Nm}(k)c{\bf N}_m(kr).
\label{jM}
\eeq
This can also be written in a more symmetrical form as
\beq
 {\bf j}=\sum_n {\bf e}_{-}j_{n,-}(k)g_{n+1}
 +{\bf e}_{+}j_{n,+}(k)g_{n-1}
 \label{jpm}
\eeq 
From  Eq.(\ref{jM}) and Eq. (\ref{jpm}) we get
$j_{n,-}=(j_{cM}-j_{cN})/2,$
$j_{n,+}=(j_{cM}+j_{cN})/2,$ $j_{cM}=(j_++j_-),$
$j_{cN}=(j_+-j_-).$

In this notation, the circuit equation becomes
\beq
{\bf Zj=E}_{ext}+E_s(\phi) {\bf e}^Z_R\delta(|w(r)|-R).
\label{bc0}
\eeq
where $Z=Z^0+\rho_0$, ${\bf e}^Z_R$ is perpendicular to the perimeter of the finite surface. 
The boundary electric field $E_s$ is the product of the normal component of the current at the surface $j_s$ and $\rho_s$, ie $E_s(\phi){\bf e}^Z_R\delta(r-R)=j_s\rho_s$ . They behave like Lagrange multipliers and will be determined from the boundary condition.

For the disk, the problem has cylindrical symmetry. Solutions of different $m$ are not coupled
to each other. Now this is not true anymore.  
We express the boundary surface field in terms of the variables $w$ in the circle as $E^s(\phi_w)=\sum_l E_s(l)\exp(il\phi_w)/(2\pi)$. As we shall see below, in the example we look at, this is a rapidly convergent expansion.

In terms of our basis states,   Eq.(\ref{bc0})  becomes: 
\begin{equation}
\sum_{k',n,i=N,M} Z_{I,k,m;k',i,n}j_{k',i,n}=E_{ext,I,k,m}+\sum_l B_I(m,l)E_s(l).
\label{ceqn}
\end{equation}

Here 
\beq
B_{I}(m,n)= \int_{\triangle} dzdz^* \exp(in\phi_w)\delta(|w|-R){\bf e^z_R\cdot I_m^*}
\label{bmn}
\eeq 
transforms the the radial component of basis function at the edge for the finite figure to that of the circle.
The integral is with respect to the variables in the finite surface.
We show below in the section on boundary electric fields that
\beq
B_{M}(n,l)=RC_n n/xJ_nd(n-l),
\label{BMN}
\eeq
$$B_{N}(n,l)=RC_{Nn}J_{n}'(kR)d(n-l).$$
where $d(n-l)= \int  d\phi_w  \exp(i(l-n)\phi)/(2|p|\pi).$ 
For the D(V) type boundary condition $B_{N(M)}=0.$

Inverting the circuit equation (\ref{ceqn}) we get 
\beq
{\bf j}= {\bf Z}^{-1} [{\bf E}_{ext}+ {\bf B E}_s]
\label{jvse}
\eeq 
The boundary condition is $j_r(R_{1}(r))=0$ for all values of $\phi$. 
From this, we finally get, 
\beq
{\bf E}_{s}=-{\bf K}^{-1}{\bf B Z}^{-1} {\bf E}_{ext}
\label{jse}
\eeq
where ${\bf K}$ is the inverse impedance projected onto the angular momentum basis of the finite surface:
\beq
K_{n,m'}=\sum_{k,I,l,m; k',J} B_I(l,n,k) Z_{I,k,l; k',J,m}^{-1}B_J(m,m',k')
\label{k}
\eeq
The resonance condition can come when the surface field is divergent, ie 
\beq
det(K)=0.
\label{res}
\eeq
It can also come when 
\beq
det(Z)=0.
\label{res1}
\eeq
Combining Ew. (\ref{jvse}) and (\ref{jse}), we get
$$
{\bf j}= [1 - {\bf Z}^{-1}{\bf B K}^{-1}{\bf B} ]{\bf Z}^{-1}{\bf E}_{ext}.
$$

Equations (\ref{jvse}), (\ref{jse}), and (\ref{res}) are the central results of this paper. What remains is
the numerical inversion of the matrix ${\bf Z}$. Usually one is interested in the low lying modes. As we see below, the diagonal elements of ${\bf Z}$ increase rapidly proportional to the square of the index. In addition, the matrix is nearly diagonal, the off-diagonal matrix elements die off rapidly. Thus one does not need many terms to get an accurate result.

\subsection{Boundary electric field}
The electric field at the boundary and the corners of a finite surface such as a triangle is one of the difficult issues that is not completely understood. In our approach with the conformal mapping, the singularity at the corner is shown to come from the singularity of the Jacobian of this mapping. 

We first determine the normal vector ${\bf e}^Z_R$ perpendicular to the boundary of the finite surface in Eq. (\ref{bc0}). We write the tangential vector as ${\bf t}
=\sum_{i=\pm} t_i{\bf e}_i.$ The boundary of the triangle is determined by the equation $w(z)w^*(z^*)=R^2$. 
Along {\bf t}, we have ${\bf t\cdot \nabla}(ww^*)=0.$ Now $\nabla={\bf e}_+\partial_z+{\bf e}_-\partial_{z^*}.$ We get $t_-w^*\partial_z w+t_+w\partial_{z^*}w^*=0.$ This can be written as
$t_-e^{-i\phi_w}p_-+t_+e^{i\phi_w}p_+=0.$ We thus get $t_{-\pm}\propto \pm e^{\pm i\phi_w}p_{\pm}.$
The normal vector ${\bf e}^z_R$ is perpendicular to {\bf t}. We thus get 
\beq
{\bf e}^z_{R}=(e^{ i\phi_w}p_{+}{\bf e}_-+e^{ -i\phi_w}p_{-}{\bf e}_+)/(2|p|).
\label{snv}
\eeq

Substituting in this expression for the normal vector in Eq. (\ref{bmn}) and after cancelling out some of the Jacobian factors we get
$$B_{M}(n,l)=R C_{Mn} n/xJ_nd(l-n)$$
$ d(m)=\int  d\phi_w  \exp(im\phi_w)/|p|/(2\pi).$ 
It is the  integrable Jacobian factor $|p|$ in the denominator of $d$ that causes the enhancement of the field at the corners.
By using the conformal transformation, this "singularity" is automatically incorporated.
Similarly
$$B_{N}(n,l)=RC_{Nn}J_{n}'(kR) d(l-n)$$

\begin{figure}[tbph]
\vspace*{0pt} \centerline{\includegraphics[angle=0,width=6cm]{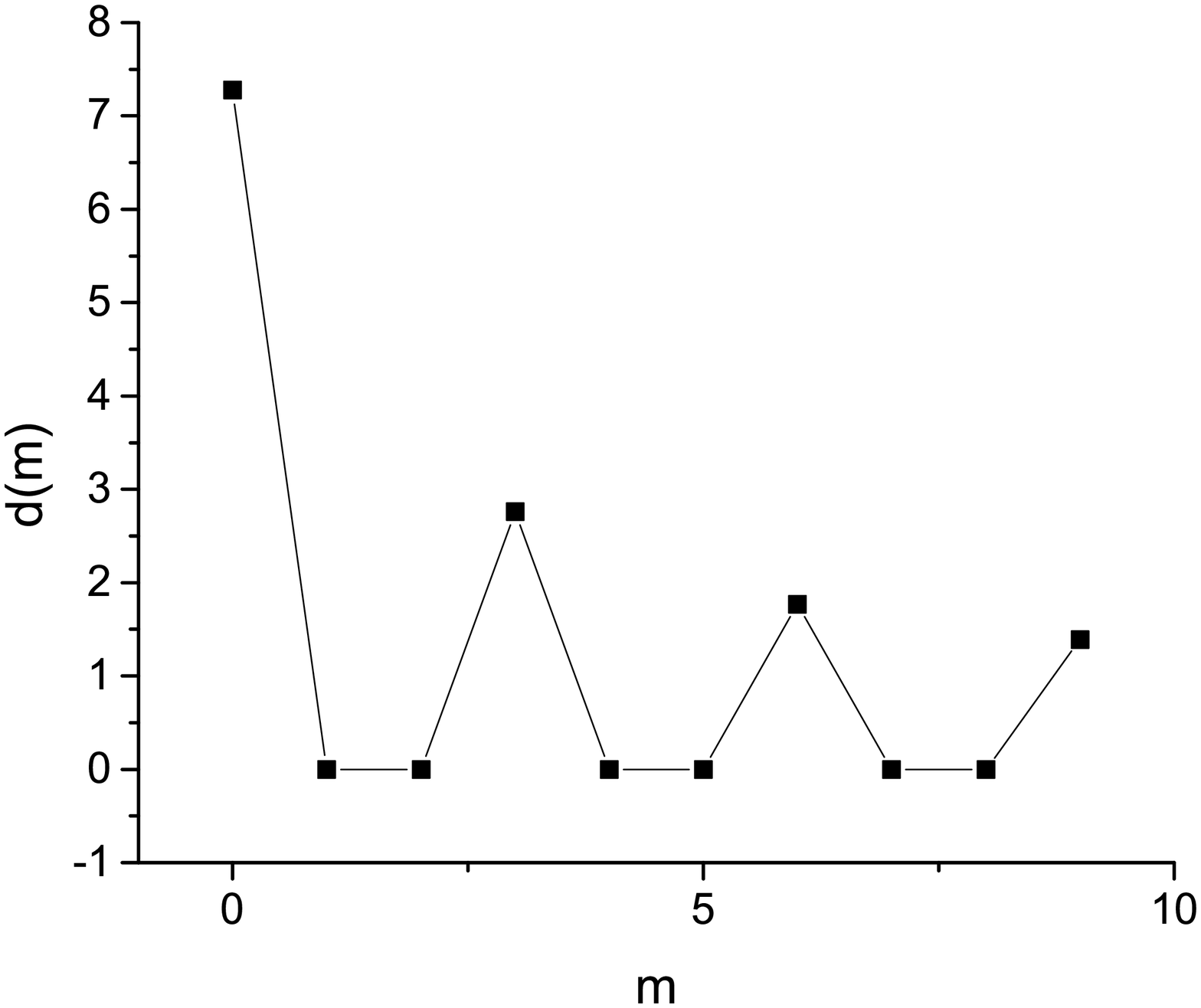}} 
\vspace*{0pt}
\caption{ d(m) }
\label{dn}
\end{figure}

For the equilateral triangle $d(m)=0$ unless $m$ is a integer multiple of 3.
It is clear from the above that $d$ is real and depends only on the absolute value of its argument.
This integral can be simply evaluated numerically. 
In this paper we shall illustrate our results with the example of an equilateral triangle. 
The value of $d(m)$ for the example of a triangle is shown in Fig. (\ref{dn})

\section{Circuit Parameters and Resonance Conditions}
From Eq. (10)
$$
L_{Xi,Yj}=\mu_0\int d{\bf r} d{\bf r'} {\bf X}_i(w(r))^*\cdot{\bf Y}_j(w'(r'))G_0(r,r').
$$

Now $\mu_0=Z_0/c$ where $c$ is the speed of light, $Z_0=(\mu_0/\epsilon_0)^{1/2}=377\Omega$ is the permittivity
of the vacuum. Thus typical magnitudes of the circuit parameters are of the order of $Z_0t$ where $t$ is the thickness of the film. 
We need the circuit parameters in terms of {\bf X} and {\bf Y}. 
For $I, J= cN, cM$ we get
$$L_{I,I}=L_{n-1,m-1}+L_{n+1,m+1},$$
$$L_{I,J}=L_{n-1,m-1}-L_{n+1,m+1}.$$
\beq
L_{n,m}=\mu_0\int d{\bf w} d{\bf w'}f_{n}^*(w) f_{m}(w')/[p_{n}(w)p_{m}^*(w')]G_0(r,r')
\label{L}
\eeq
In comparison to the expression for the circle, there are now additional factors 1/p which becomes large at the corners. The integral is non-divergent.
The numerical evaluation of different elements can be trivially calculated on different processors of a parallel computer, as that for different indicies are independent of each other.
We next discuss its symmetry properties for some specific cases.

We write $L=L'+iL''$ with $L'$, $L''$ coming from the real and imaginary parts of the Green's function:
 $L'_{n,m}=\mu_0\int d{\bf w} d{\bf w}' F,$
 $L''_{n,m}=\mu_0\int d{\bf w} d{\bf w}' G.$
$F=[p_n(w)p_m(w')^*]^{-1}f_n(w)^*f_m(w')\cos(k_0|r-r'|)/|r(w)-r(w')|$, $G=[p_n(w)p_m(w')^*]^{-1}f_n(w)^*f_m(w')\sin(k_0|r-r'|)/|r(w)-r(w')|.$ 
For systems with reflection symmetry both $L'$ and $L''$ are real. 
This can be seen as follows: If we make the change of variable $\phi$ to $-\phi$ and  $\phi'$ to $-\phi'$, the integrand becomes its complex conjugate. The integral  contain contributions from all possible values of the angle and is thus the sum of contributions from both signs of the angles. Thus $L'$ $L''$ are real.
Similarly 
$$L^{\prime}_{n,m}=L'_{n,m},$$
$$L''_{m,n}=L''_{n,m}.$$

We next look at the capacitance given by
$$
1/C_{Xi,Yj}=-\int d{\bf r} d{\bf r'}\nabla\cdot c{\bf X}_i(r)^*\nabla '\cdot c{\bf Y}_j(r')G_0(r,r')/\epsilon_0
$$ 

From Eq. (\ref{div}) the only nonzero components of the
capacitance are given by
\beq
1/C_{Ni,Nj}=-k_ik_j\int d{\bf w} d{\bf w'}f_n(k_iw)^*f_n(k_jw')G_0(r,r')/\epsilon_0
\label{C}
\eeq

Whereas the inductive term contain additional factors of 1/p, no such factor appears in the expression for 1/C. 
The other components are zero: $$(1/C)_{X,j; i,M}=0.$$ $1/C(k,k') $ increases rapidly as $k$, $k'$ is increased. We are interested in the inverse of the impedance matrix. This makes the problem of inverting the impedance matrix rapidly convergent and is one the the simplifying feature of the present approach. 

We next discuss results for the example of a equilateral triangle.
The conformal mapping between a point z inside a regular polygon of K sides into the point w inside a disk is given by
$$z =  RwF(1/K; 2/K; 1 + 1/K;w^K)
$$
where $F(a,b,c;z)$ is the confluent hypergeometric function.
The points $w_{k} = \exp (i2\pi k/K)$ on the unit circle is  mapped into the polygon
corners :
$z_{k} = RC(K) \exp i( 2\pi k/K)$ 
where
\beq
C(K) =\Gamma(1 + 1/K)\Gamma(1 - 2/K)/\Gamma(1 - 1/K).
\label{ck}
\eeq
The details of this is recapitulated in Appendix 1. 
Because of the three fold rotation symmetry of the surface,
$L_{m,n}$ is nonzero only if $m=n+3i$ for integer $i$.

\begin{figure}[tbph]
\vspace*{0pt} \centerline{\includegraphics[angle=0,width=6cm]{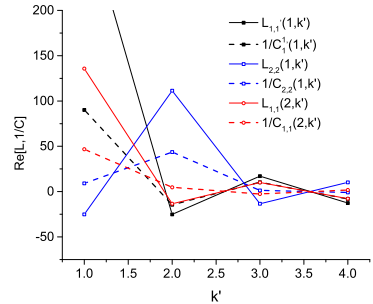}} 
\vspace*{0pt}
\caption{ Dependence L (in units of $\mu_0/[4\pi\sqrt{3}C(3)]$)
$1/C(3)^4$) 
and 1/C (in units of $\sqrt{3}C(3)/[4\pi\epsilon_0]$
vs k' in units of $1/[\sqrt{3}C(3)]$. }
\label{ldck}
\end{figure}

\begin{figure}[tbph]
\vspace*{0pt} \centerline{\includegraphics[angle=0,width=6cm]{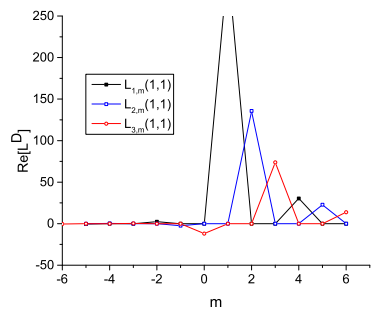}} 
\vspace*{0pt}
\caption{ Circuit parameters $L_{XX}(i,m)$ for the D type boundary condition in units of $\mu_0/[4\pi\sqrt{3}C(3)]$.}
\label{lim}
\end{figure}

\begin{figure}[tbph]
\vspace*{0pt} \centerline{\includegraphics[angle=0,width=6cm]{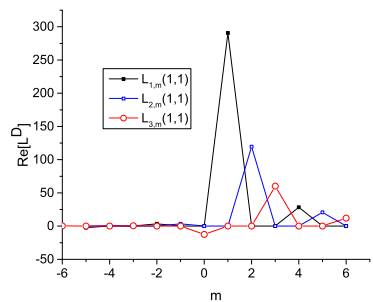}} 
\vspace*{0pt}
\caption{ The off-diagonal inductance  $L_{XY}(i,m)$  for the D type boundary condition in units $\mu_0/[4\pi\sqrt{3}C(3)]$.}
\label{lxyim}
\end{figure}

\begin{figure}[tbph]
\vspace*{0pt} \centerline{\includegraphics[angle=0,width=6cm]{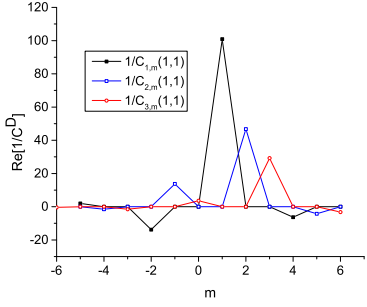}} 
\vspace*{0pt}
\caption{ Circuit parameters $1/C_{NN}(i,m)$  for the D type boundary condition in units of $\sqrt{3}C(3)/[4\pi\epsilon_0]$.}
\label{cim}
\end{figure}

We have calculated the circuit parameters by directly performing the integrals numerically. As we emphasized, the numerical computation can be trivially vectorized; different components can be computed on different processors. As we have explained before\cite{book,disk} we expect the impedance matrix $X(k,k')$
[$X_{m,m'}$] to be nearly diagonal. This is indeed true, as is illustrated in our Fig. (\ref{ldck}) [Fig.(\ref{lim})] where we show the k [m] dependence of the matrix elements 
given in Eq. (\ref{L},\ref{C}).
The circuit elements also decreases as the angular momentum index is increased. This is illustrated in fig. (\ref{lim}) and (\ref{lxyim}) where we show some of the diagonal and off-diagonal components of inductance matrix. Similar result is observed for the capacitance matrix, as is shown in Fig. (\ref{cim}). 

\section{Numerical results for the resonance behaviour of the equilateral triangle}

We have applied our method to the case of an equilateral triangle with edges of length $a=1$. 
For material such as Cu of a thickness of $t=3 nm$ (the skin depth at infrared frequencies), 
the effective intrinsic resistance
$\rho_0/t\approx 3 \Omega.$ This is much less than the impedance, which is of the order of $Z_0=377 \Omega$.
Using the circuit parameters discussed above, we have solved the resonance condition in Eq.(\ref{res}).
This is achieved by calculating the eigenvalues of the matrix $K$ for a mesh of frequencies.
The resonance condition is determined by the condition that the real part of one of the eigenvalue
changes sign. The imaginary parts of the eigenvalues, due to the intrinsic and the radiative resistances, are much smaller. Including the resistances in our calculation 
changes the resonance frequencies by less than 1 per cent.
We have included the lowest four $k$'s with m ranging from -6 to +6 and have checked our results to make sure that convergence is achieved. Because of the three fold rotation symmetry of the equilateral triangle, a resonance is characterized by angular momenta $m+3n$ with different m and all possible values of n. We found a variety of resonances that range from dark to grey to bright. Some of these are degenerate. We first discuss the doubly degenerate lowest resonances.

\subsection{Lowest Resonances}
\begin{figure}[tbph]
\vspace*{0pt} \centerline{\includegraphics[angle=0,width=6cm]{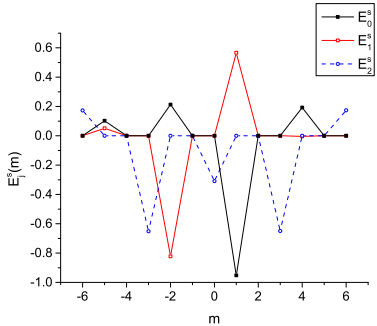}} 
\vspace*{0pt}
\caption{ The angular momentum components of the "bare" surface field for the lowest three resonances. For the case when there are degenerate resonances, only one such is shown. 
}
\label{esf}
\end{figure}

Our lowest resonance comes from the D type boundary condition. It is doubly degenerate. The information on the degeneracy is difficult to extract from a calculation of the scattering cross section but is trivial in the
current approach. The resonance frequency is at   $\omega_r=\pm 2.4.$ 
The boundary field components $E_s(m)$ of one of the 
degenerate resonance is shown as the black solid line in Fig.(\ref{esf}).
Because of the three fold rotation symmetry of the system, angular momentum components that differ 
from each other by factors of three are coupled to each other. The two degenerate modes 
correspond to linear combinations of ( m=+1, -2, -5...) and (m=-1, 2,
5..). In terms of this surface field, the eigenstate $|e>$ is very simple. 
The resonance is dominated by a single angular momentum component in terms of the basis functions of the circle. The other degenerate resonance corresponds to one with $E'_{s}(m)=E_{s}(-m).$

\begin{figure}[tbph]
\vspace*{0pt} \centerline{\includegraphics[angle=0,width=10cm]{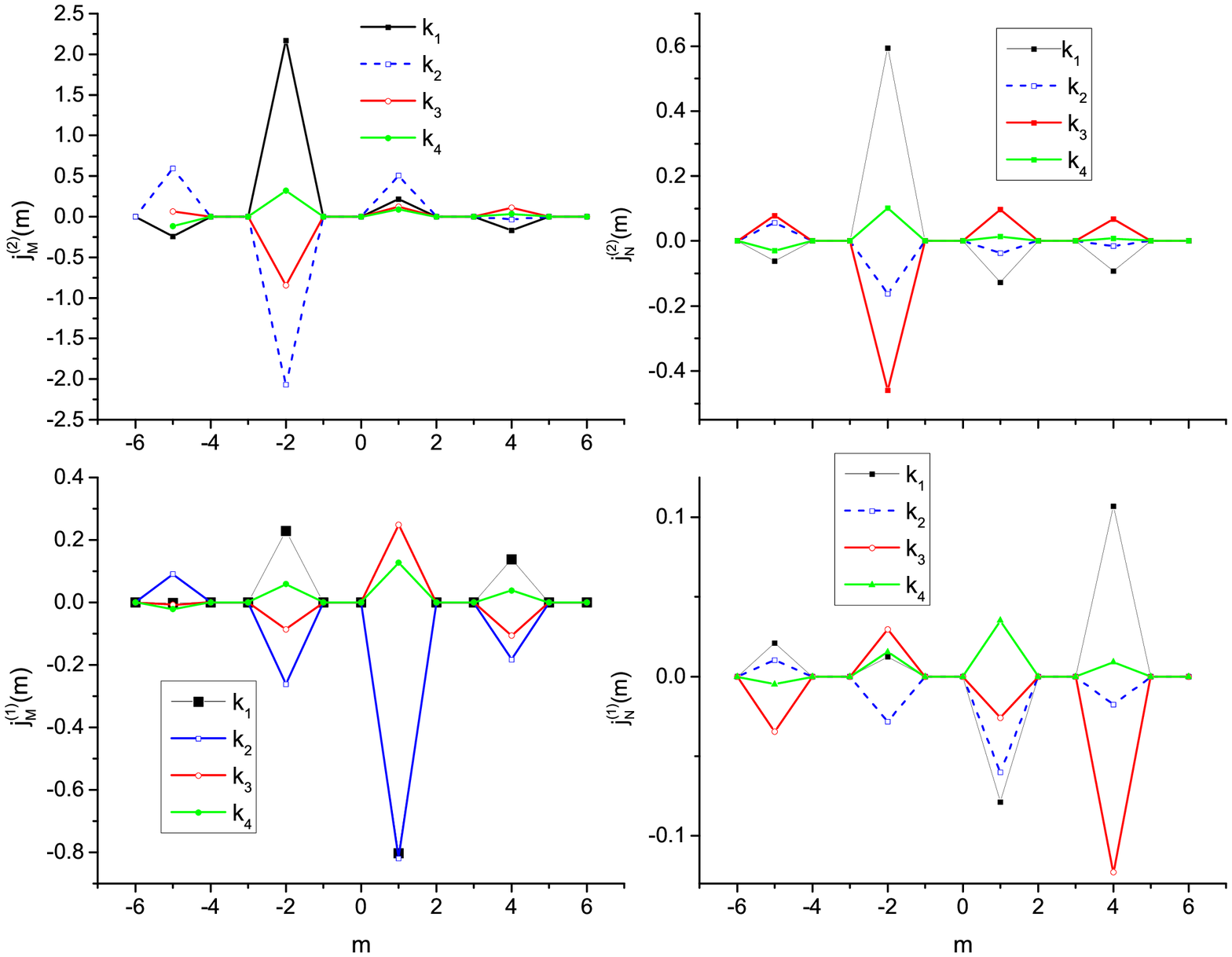}} 
\vspace*{0pt}
\caption{The angular momentum components of the resonances in the triangle for the lowest two resonances. The {\bf M} ({\bf N}) components are left (right). The lowest resonance is at the bottom.}
\label{efmk}
\end{figure}

\begin{figure}[tbph]
\vspace*{0pt} \centerline{\includegraphics[angle=0,width=10cm]{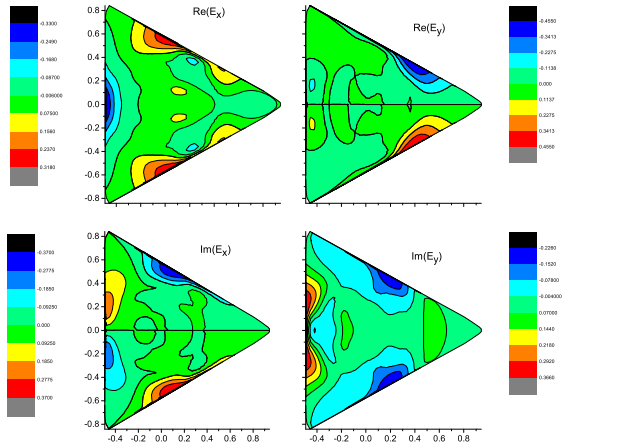}} 
\vspace*{0pt}
\caption{ The real and imaginary part of the resonance field in the x  and the y direction. The x (y) component is on the left (right). The real (imaginary) part is at the top (bottom). The normalization is arbitrary. The length units are such that the distance from the center to tr corner is 1.}
\label{evec}
\end{figure}

This apparently simple $E_{s}(m)$ corresponds to a rich pattern of electric currents inside the triangle.
The components of the eigenstate 
in terms of the basis function of the triangle is shown in Fig. (\ref{efmk}). The vertical
scale on the right for the N components is smaller than that on the left for the M components.
The real and imaginary part of the resonance field (current) in the x and the y direction can be calculated from Eq. (\ref{jvse}) and is shown in Fig.\ref{evec}.

\begin{figure}[tbph]
\vspace*{0pt} \centerline{\includegraphics[angle=0,width=6cm]{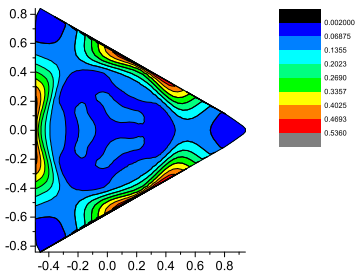}} 
\vspace*{0pt}
\caption{ The magnitude of the resonance field.  The normalization is arbitrary. The length units are such that the distance from the center to tr corner is 1.}
\label{emag}
\end{figure}

The magnitude of the currents are largest near the edges. 
The three fold rotation symmetry of the triangle is not explicit from the above figures.
The magnitude of the resonance field is  shown in 
Fig. (\ref{emag}). As can be seen, it is indeed three-fold symmetric.

We have also estimated the electric susceptibility of this system. We approximate $K^{-1}$ in Eq. (\ref{k}) by
$K^{-1}\approx |e><e Z_0^2/\rho_0$ where $Z_0$ is the permittivity of the vacuum, $\rho_0$ is the effective intrinsic resistivity of the metal.
The electric susceptibility of the resonance is then given by
\beq
\chi \approx \int_{\triangle} P <{\bf Z}^{-1}{\bf B }|v><v|{\bf B} ]{\bf Z}^{-1}>/\rho_0.
\eeq
Here $P$ is the electric dipole moment given in Eq. (\ref{d}). The magnitude of the electric  susceptibility is
$|\chi| = 0.021 a/(c\rho_0)$ 
where $a$ is the length of a side of the triangle.

\begin{figure}[tbph]
\vspace*{0pt} \centerline{\includegraphics[angle=0,width=6cm]{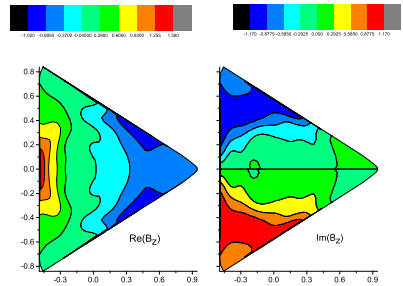}} 
\vspace*{0pt}
\caption{ The magnetic resonance field. The real (imaginary) part is on the left (right). The normalization is arbitrary. The length units are such that the distance from the center to tr corner is 1.}
\label{bz1}
\end{figure}

This resonance has no global magnetic moment because it does not contain any zero angular momentum component. It does contain local magnetic moments, however.  
The corresponding perpendicular component of the magnetic field $B_z$ (vorticity) of this resonant state is shown in Fig. \ref{bz1}.

\subsubsection{Results from other numerical methods}
\begin{figure}[tbph]
\vspace*{0pt} \centerline{\includegraphics[angle=0,width=6cm]{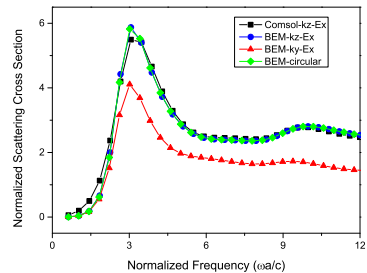}} 
\vspace*{0pt}
\caption{Scattering cross section}
\label{scs}
\end{figure}

 To substantiate our results, we have also calculated the scattering cross section of a triangle 
using different numerical approaches such as the boundary element method (BEM)\cite{BEM}, 
a Green's-function-based full-wave integral equation method for simulating electromagnetic wave propagation and scattering problems. Similar result is also obtained with the commercial finite-element method package
COMSOL\cite{comsol}.
Our results for the scattering cross section $S$ as a function of frequency 
is shown in Fig. \ref{scs}. $S$ peaks at a frequency $\omega_1 a/c\approx 3$.
For these numerical results, there is a background contribution that is increasing with frequency.
More precisely, the scattering cross section $S$ is a sum of a background term $S_b$ and a resonance contribution $S_r.$ 
For example, in Eq. (\ref{jvse}), there are two terms on the right hand side, the  resonance is from the second term and comes from the divergence of the  surface field $E^s$. The first term on the right hand side provides for the  background term.
In general, a response function is from a sum over many eigenstates with  different denominators $\omega-E_i$. Only a few of them are resonant with  $E_i=\omega$, The other contributions with $E_j$ not equal to $\omega$  provides for the background term.

The maximum of $S$ occurs when $dS/d\omega=0$. This is equivalent to $dS_r/d\omega=-dS_b/d\omega$ and differs from the condition for the maximum of $S_r$.  
After the background contribution is subtracted off, the peak position for the resonance term will be shifted down.  A crude estimate suggests that $S_b\approx (\omega a/c-1)$, $S_r\approx 5 \exp -[(\omega-\omega_r) a/(2c)]^2$. From this we obtain $\omega_r a/c=2.5.$ So we believe the results for the resonance frequency in this paper is in reasonable agreement with the above.

\begin{figure}[tbph]
\vspace*{0pt} \centerline{\includegraphics[angle=0,width=6cm]{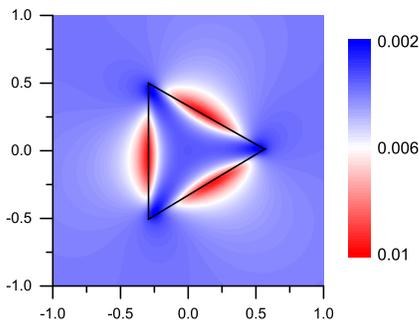}} 
\vspace*{0pt}
\caption{Magnitude the component of the magnetic field $H$ parallel to the plane directly above the triangle.}
\label{Hnorm}
\end{figure}

The magnitude of the component of the magnetic field $H$ parallel to the plane directly above the triangle is shown in Fig(\ref{Hnorm}). This quantity is proportional to the magnitude of the current in the plane.
As can be seen, it agrees with 
the result shown in Fig. (\ref{emag}). 


\subsection{Grey Resonances}
The next resonances are also doubly degenerate and occur at frequencies   $\omega_r=\pm 5.16$. 
The boundary field components $E_s(m)$ of one of the 
degenerate resonance is shown in Fig.(\ref{esf}).
The two degenerate modes again
correspond to linear combinations of ( m=+1, -2, -5...) and (m=-1, 2,
5..). 
For the previous resonance, it is dominated by the $|m|=1$ contribution. For this one  
the $|m|=2$ contribution is more important. 
The components of the eigenstate 
in terms of the basis function of the triangle is also shown in Fig. (\ref{efmk}). Again 
the $|m|=2$ contribution is more important. Because electric dipole comes from the $|m|=1$ components
we expect it to be smaller. We find that
the electric susceptibility $|\chi|=0.0029 a/(c\rho_0)$, 
an order of magnitude smaller than that for the previous resonance. Again, the magnetic dipole moment is zero because it does not contain the zero angular momentum component. 

\subsection{Pure magnetic resonance}
The next resonance is  nondegenerate and occurs at a frequency  $\omega_r=\pm 5.7$ 
The boundary field components $E_s(m)$ of  this resonance is shown in Fig. (\ref{esf}).
This mode corresponds to linear combinations of ( m=+3, 0, -3, ...).
The components of the eigenstate 
in terms of the basis function of the triangle is also shown in Fig. (\ref{efmk}). 
This state does not have any $|m|=1$ terms and thus no electric dipole moment.
It contains a $m=0$ term and thus have a finite magnetic dipole moment .




In summary, we have exploited conformal mapping to generate a convenient set of orthonormal basis functions so that the response to an external electromagnetic field in this basis is rapidly convergent.
This provided a simple way to study different physics issues  and provided new insight that is not easily obtained. This included  the issue of vorticity and eddy current, and the nature of the divergent electric field at the boundaries and at corners. In particular the nature of the resonance can be directly addressed. 

We study an example of an equilateral triangle, we found that the lowest resonance is doubly degenerate and magnetically dark.
Good agreement is found with results obtained with traditional numerical techniques.
We hope our results can be applied to offer new understanding for other traditional structures. 

STC thanks the hospitality of the Physics Dept. and the Institute for Advanced Studies of the Hong Kong University of Science and Technology where this work is finished. He also thanks T. Ishihara for discussions of work carried out together with Q. Bai; Ed Nowak for a discussion of noise in different materials.
  

\section{Appendix 1: Conformal Mapping}
We recapitulate the conformal mapping between a circle and a triangle
with the Schwartz-Christoffel (SC) transformation for a polygon with verticies $w_k$ and angles $\alpha_k/\pi$.
The explicit formula for the SC transformation is
$$z = S(w) = A\int^w dw'(w' -w_1)^{-\alpha_1/\pi} ... (w' -w_k)^{-\alpha_k/\pi}$$

For the regular K-polygon, $z = S(w) = A\int_0^w dw'G, $ 
$G=\Pi_k[1-w'\exp(-i 2\pi k/K)]^{-2/K}=[1-(w')^K ]^{-2/K}.$
After integrating the infinite series for G, one gets
$$z = S(w) 
=A\sum_{n=0}^{\infty} (2/K+n-1)!w^{Kn+1}/[(Kn+1)n! ].$$
Note also, that 
$dz/dw=G(w)=(1-w^K )^{-2/K}.$

The mapping can also be written as:
$$z =  RwF(1/K; 2/K; 1 + 1/K;w^K )
$$
where $F(a,b,c;z)$ is the confluent hypergeometric function, R is a parameters.
The points $w_{k} = \exp i(2\pi k/K)$ on the unit circle is  mapped into the polygon
corners :
$z_{k} = RC(K) exp i( 2\pi k/K)$ 
where
\beq
C(K) =\Gamma(1 + 1/K)\Gamma(1 - 2/K)/\Gamma(1 - 1/K).
\eeq

If $R=1,$ the triangle vertex is at $r=C(3)=1.7666$
The edge length is $a=2C(3)\cos \pi/6=\sqrt{3}C(3).$
In units with $a=1$, the resonance frequency $\omega^2\propto 1/(LC)\propto 1/R^2$is increased by a factor
of $a^2$.


\section{Appendix 2: Derivative}
We need the derivative with respect to the polygon coordinates (Note that from the Riemann Cauchy relationship, $\partial_z(w^*)=0$ ). We use the notation $\nabla'=\nabla_w,$ $\nabla=\nabla_r$ 
for the derivatives with respect to the coordinates of the circle and the triangle, respectively. We have 
\beq
\nabla={\bf e}_+\partial_z+{\bf e}_-\partial_{z^*}
\eeq
where $\partial_z=(\partial_x-i\partial_y)/2.$ 
Recall that $g_{n\pm 1}=p_{\pm}f_{n\pm 1}.$ 
Because $\partial_w p_+=0,$ we get 
\begin{equation}
\nabla\cdot c{\bf M}=(p_+\partial_w  f_{n+1}\partial w/\partial z+ p_-\partial w^*/\partial z^* \partial_{w^*} f_{n-1}).
\label{divm}
\end{equation} 
It is straightforward to show that
\beq \partial_w f_{n+1}
=k J_{n}e^{in\phi}/2,\ 
\partial_{w^*} f_{n-1}
=-k J_{n}e^{in\phi}/2.
\label{der}
\eeq
From Eq. (\ref{divm}) 
$\nabla\cdot c{\bf M}=0.$ {\bf M} has zero divergence. A current distribution described by this state is of magnetic character and does not generate a finite charge density. 

If we had used a diffferent choice of p, we would have an additional term
$m=(f_{n+1}\partial_w p_+ \partial w/\partial z+ f_{n-1}\partial w^*/\partial z^* \partial_{w^*} p_-).$
Similarly, 
\begin{equation}
\nabla \cdot c{\bf N}
= -k f_n|p|^2.
\end{equation} 

We next look at the curl. Let ${\bf E}=\sum_{i=\pm}E_ie_i$
We get
$$({\bf \nabla\times E})_z=2i[\partial_{z^*} E_+ -\partial_zE_-].$$
For a general electric field
$$(\nabla\times E)_z
=i\sum_n|(dw/dz)|^2(c_{+n}\partial_{w^*}f_{n-1}-c_{-n}\partial_{w^*}f_{n+1})
$$
$$
=ik\sum_n|(dw/dz)|^2(c_{+n}+c_{-n})f_{n}).
$$
For {\bf M} ({\bf N}) $c_+ = +(-) c_-=1/2$. 
Thus 
\beq
(\nabla\times M)_{nz}=ik|(dw/dz)|^2f_{n},
\eeq
$(\nabla\times N)_z=0.$
The states  {\bf M}  possess finite z angular momenta and correspond to states of finite vorticities.

%


\begin{references}
\bibitem{bookm} David Gu and S. T. Yau, ''Computational conformal geometry'', International Press, Boston, (2008).
\bibitem{book} S. T. Chui and Lei Zhou, " Electromagnetic behaviour of metallic wire structures'',
Springer, (2013).
\bibitem{disk} S. T. Chui, J. J. Du and S. T. Yau, Phys. Rev. E 90, 053202 (2014).
\bibitem{zhan} Multiple scattering of metallic wire structures
T. R. Zhan, S. T. Chui, and Z. F. Lin
Journal of Applied Physics 118, 163104 (2015); doi: 10.1063/1.4934492
\bibitem{detail2} See, for example, chap. 15 in H. C. van de Hulst,
''Light Scattering by small particles'', Dover (New York).
\bibitem{Jackson} See, for example,  Eq. (10.14) in ''Classical Electrodynamics'' by J. D. Jackson,
Wiley, 1st Ed., (1962).
\bibitem{BEM}
See, for example, S. B. Wang, H. H. Zheng, J. J. Xiao and C. T. Chan, 
Int. J. Comp. Mat. Sci. Eng. 01, 1250038 (2012).
\bibitem{comsol} Information can be obtained from the web site www.comsol.com.
\end{references}
\end{document}